\documentclass[fleqn,twoside]{article}
\usepackage{amsmath}
\usepackage{amssymb}
\usepackage[headings]{espcrc2}
\usepackage{url}

\readRCS
$Id: espcrc2.tex,v 1.2 2004/02/24 11:22:11 spepping Exp $
\usepackage{graphicx}
\usepackage[figuresright]{rotating}

\newcommand{\as}{\alpha_s}

\newcommand{\tw}{\textwidth}
\newcommand{\cF}{{\cal F}}

\newcommand{\cE}{{\cal E}}
\newcommand{\cR}{{\cal R}}

\newcommand{\ie}{\emph{i.e.}\ }
\newcommand{\eg}{\emph{e.g.}\ }
\newcommand{\cnf}{\emph{cf.}\ }

\newcommand{\ee}{e^+e^-}

\newcommand{\order}[1]{{\cal O}\left(#1\right)}

\title{Automated Resummation and Hadron Collider Event Shapes%
  \thanks{Talk presented at the 7th International Symposium on
    Radiative Corrections (RADCOR 2005), Shonan Village, Japan, Oct.
    2-7, 2005 } 
}

\author{Gavin P.~Salam\address{LPTHE, Universit\'e Pierre et Marie
    Curie -- Paris 6, Universit\'e Denis Diderot -- Paris 7, CNRS UMR
    7589, 75252 Paris 75005, France}}
       
\runtitle{Automated Resummation and Hadron Collider Event Shapes}
\runauthor{Gavin P.~Salam}

\begin{document}

\begin{abstract}
  This writeup gives an introduction to the theoretical understanding
  that lies behind automated resummation. It then discusses its
  applications to hadron-collider event shapes.\vspace{1pc}
\end{abstract}

% typeset front matter (including abstract)
\maketitle

%----------------------------------------------------------------------
%\section{INTRODUCTION}

The all-order resummation of logarithmically enhanced terms is
essential for the detailed study of final-state observables like event
shapes and jet rates (see for example~\cite{DSReview}). The
differential distribution for an event shape $V$ to have value $v$ can
be written as
\begin{equation}
  \label{eq:diffdist}
  \frac{1}{\sigma}\frac{d\sigma}{dv} \equiv \Sigma'(v) = \as f_1(v) +
  \as^2 f_2(v) + \ldots
\end{equation}
where the leading (LO) and next-to-leading (NLO) order coefficients
$f_1$ and $f_2$ can be calculated with a program such as
NLOJET~\cite{NLOJET}. At small $v$, for any observable
sensitive to soft collinear radiation, the corresponding integrated
distribution $\Sigma(v)$ has double-logarithmic enhancements at all
orders
\begin{subequations}
  \label{eq:doubleLogs}
\begin{align}
  \Sigma(v) &\simeq \sum_{m} \sum_{n=0}^{{2m}} H_{mn} \as^m \ln^n \frac1v
   \qquad (v\to 0)\\
  &= \underbrace{h_1(\as L^2)}_{LL_{{\Sigma}}} \,+\,
  \underbrace{\sqrt{\as}\, h_2(\as L^2)}_{NLL_{{\Sigma}}} \,+ \ldots
  \label{eq:doubleLogsB}
\end{align}
\end{subequations}
where $H_{mn}$ are numerical coefficients and we have introduced the
shorthand $L \equiv \ln 1/v$. The double logarithms are associated
with non-cancellation between soft and collinear divergences in real
and virtual graph, since limiting $v$ suppresses real radiation but
not the corresponding virtual terms. 

When $v$ is small ($\ln v \sim \as^{-1/2}$) all terms $\as^m L^{2m}$
become of the same magnitude ($\sim 1$) and reliable calculations for
$\Sigma(v)$ can only be obtained by resumming these terms to all
orders, giving the leading logarithmic (LL) function $h_1(\as L^2)$.
One can systematically reorganise the whole perturbative series in
terms of the logarithmic enhancements, eq.~(\ref{eq:doubleLogsB}),
with next-to-leading logarithmic (NLL) terms, $\as^m L^{2m-1}$
resummed into a function $\sqrt{\as} h_2(\as L^2)$, etc., such that
each of the $h_n$ is of order $1$.

For yet smaller $v$, $L \gg \as^{-1/2}$, even the reorganisation of
eq.~(\ref{eq:doubleLogs}) breaks down, since the functions $h_n$ can
themselves become large as their argument, $\as L^2$, grows beyond
$1$. It turns out however that quite often eq.~(\ref{eq:doubleLogs})
can be written as the exponential of a much less divergent series,
\begin{subequations}
  \label{eq:expForm}
\begin{align}
  \label{eq:expFormA}
  \ln \Sigma(v) &\simeq \sum_{m} \sum_{n=0}^{{m+1}} G_{mn} \as^m L^n
  \qquad (v\to 0) \\
  &= \underbrace{L g_1(\as L)}_{LL_{{\ln \Sigma}}} \,+
  \,\underbrace{g_2(\as L)}_{NLL_{{\ln \Sigma}}} \,+ \ldots\,,
  \label{eq:expFormB}
\end{align}
\end{subequations}
where the key feature of eq.~(\ref{eq:expFormA}) is that the index $n$
now runs only up to $m+1$, \ie all the double logarithms of $\Sigma$
arise from the exponentiation of a single $\as L^2$ term in $\ln
\Sigma$. This series can be also reorganised systematically,
eq.~(\ref{eq:expFormB}) and now one redefines the LL terms to be $L
g_1(\as L)$ and the NLL to be $g_2(\as L)$, etc., where the $g_n(\as
L)$ have the property that they are all of order $1$ as long as $L
\sim \as^{-1}$. This means that the reorganised perturbative hierarchy
remains stable to much smaller values of $v$, than in
eq.~(\ref{eq:doubleLogs}); furthermore the NLL terms are more
suppressed with respect to the LL terms (by $1/L\sim \as$) than was
the case in the classification of eq.~(\ref{eq:doubleLogsB}).  

\begin{figure*}
  \centering
  \includegraphics[width=0.8\textwidth]{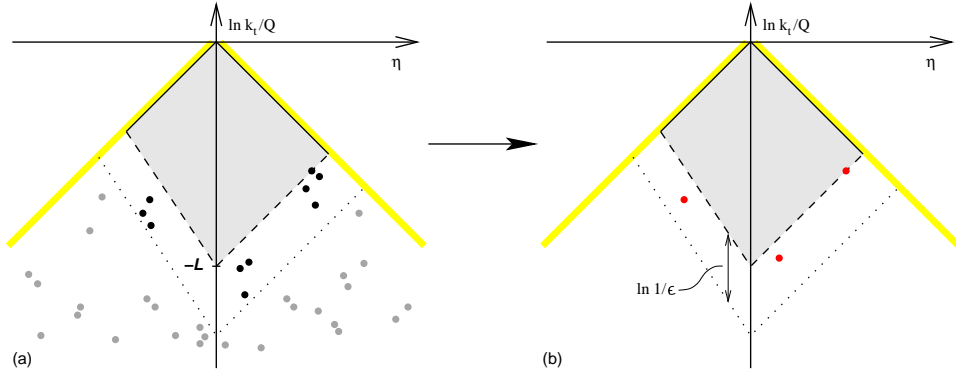}\vspace{-0.5cm}
  \caption{Representation of phase space in $\ee \to 2$~jets;
    emissions (points) in the right (left) hand half-plane are
    collinear to the right (left) hand jet. For further details, see text.}
  \label{fig:clusters}
\end{figure*}

%----------------------------------------------------------------------
%\section{GENERALISED RESUMMATION}

In this writeup it is the more ambitious, exponentiated, kind of
resummation that will be discussed. Let us examine the kinematic plane
in $\ee$, fig.~\ref{fig:clusters}, represented in terms of the
transverse momentum $k_t$ and rapidity ($\eta =
\frac12\frac{E+k_z}{E-k_z}$) of an emission $k$, as measured with
respect to the Born $q\bar q$ axis. The thick yellow lines represent
the hard-collinear kinematic limit for emissions. If there is just a
single emission, one calculates the $\order{\as}$ term of the
logarithmically enhanced part of $\Sigma(v)$ by identifying the grey
shaded area in which the observable's value\footnote{For
  compactness we write the observable as a function just of the soft
  and collinear emission momenta, though there is also an implicit
  dependence on the hard momenta.} $V(k)$ is larger than $v$.  In the
single-emission approximation this region is forbidden to real
radiation, but not to virtual corrections, and the resulting
non-cancellation leads to a double logarithmic contribution $\as L^2$
to $\Sigma(V)$.

In general, when considering multiple emissions, the boundary
$V(k_1,\ldots,k_n) < v$ will be a complex multi-dimensional function
of the $k_i$. For many observables there are several simplifications
that can be carried out. These will look quite similar to those, based
on infrared and collinear (IRC) safety, that are used to justify
fixed-order calculations. We use as a reference scale the lower
boundary (dashed line) of the shaded region in
fig.~\ref{fig:clusters}a, corresponding to a value $\sim v = e^{-L}$
of the observable for a single emission. Regardless of the set of
emissions that is present, we require that all emissions much softer
than the scale of the boundary should modify the observable by much
less than $v$. These emissions (in grey, below the dotted line) can
therefore be neglected, since they will cancel fully against
corresponding virtual terms. Then, only the remaining, black
emissions, between the dashed and dotted lines, will contribute
significantly to the observable. These are clustered in rapidity since
there is a collinear divergence for the splitting of an emitted
parton. However, most common observables have the property that they
are insensitive to this collinear splitting, so that each cluster of
black emissions can be replaced by a single red `primary' emission, as
shown in fig.~\ref{fig:clusters}b.

For the next stage, it is necessary to exploit another property common
to most observables, namely that
\begin{equation}
  \label{eq:simsimsim}
  V(k_1,\ldots,k_n) \sim \max[V(k_1),\ldots,V(k_n)]\,,
\end{equation}
so that if $V(k_1,\ldots,k_n) < v$ then all $V(k_i) \lesssim v$. This
means that the shaded region will contain no emissions (except
potentially near its lower edge) implying a resummation of purely
virtual corrections there. This leads to the exponentiated double
logarithmic contribution $L g_1(\as L)$.

What remains to be calculated is the correction coming from real
emissions, confined to the band (approximately) between the shaded and
dotted lines. As long as the observable is such that one can ensure
that this band has a width $\ln 1/\epsilon$ that is not parametrically
large (\ie independent of $L$), then the band will give at most
single-logarithmic contributions $\as^n L^n$, \ie it will contribute
only to the $g_2(\as L)$ function, not to $g_1$.  Because the density
of primary emissions per unit rapidity in the band, $\sim \as \ln
1/\epsilon$ will be low, emissions will be widely separated in
rapidity, allowing one to make use of the property of angular ordering
whereby the full matrix element for multiple emission reduces to that
for independent emission~\cite{CTTW}, simplifying the calculation of
$g_2$. Note that this only works for continuously global observables
\cite{DasSalGlob,DasSalDIS}.

To aid the mathematical treatment of the above procedures it is
convenient to parametrise the observable's dependence on a single
emission close to hard leg $\ell$ (in $\ee \to 2$~jets, $\ell=1,2$) as
\begin{equation}
  \label{eq:Vparam}
  V(k) = d_\ell g_\ell(\phi) \left(\frac{k_t}{Q}\right)^a e^{-b_\ell \eta}\,,
\end{equation}
with $k_t$ and $\eta$ now measured with respect to leg $\ell$, and
$a$, $b_\ell$ and $d_\ell$ numerical constants and $g_\ell(\phi)$ a
function of the azimuthal angle $\phi$, all of them
characteristic of the observable and relevant to determining the
position of the dashed line in fig.\ref{fig:clusters} for a given $v$
(where $\phi$ is not represented). One also defines $\bar d_\ell =
d_\ell \exp\, [\,\int_0^{2\pi}\! \frac{d\phi}{2\pi} \ln
g_\ell(\phi)]$.

It will be useful to express momenta in terms of the effect they
have on the observable, through functions $\kappa_i(\zeta)$ that
satisfy
\begin{equation}
  \label{eq:kappai}
  V(\kappa_i(\zeta)) = \zeta\,.
\end{equation}
This condition alone is not enough to fully specify $\kappa_i(\zeta)$
--- additionally one needs to know the leg $\ell_i$ to which it is
collinear, its azimuthal angle $\phi_i$, and how its rapidity depends
on $\zeta$ --- for concreteness here we will take this to be
$\eta_i(\zeta) = \xi_i (\ln 1/\zeta)/(a+b_\ell)$ so that taking
$\xi_i$ in the range $0 < \xi_i < 1$, $\kappa_i(\zeta)$ will span from
the large-angle soft region to the hard-collinear region (regardless
of $\zeta$). 

Part of the logic of this scaling is that the emission pattern $\sim
\as \frac{dk_t}{k_t}d\eta$ becomes proportional to $\as L \cdot d\ln
\zeta d\xi$, with the single logarithmic factor $\as L$ factoring out
explicitly. It also helps compactify conditions such as
eq.~(\ref{eq:simsimsim}), which becomes
\begin{equation}
  \label{eq:simsimTrans}
  V(\kappa_1(\zeta_1),\ldots,\kappa_n(\zeta_n)) \sim
  \max(\zeta_1,\ldots \zeta_n)\,.
\end{equation}
Actually ``$\sim$'' is difficult to encode directly in a computer, as
will be needed for automation. Here the meaning of ``$\sim$'' is that the
ratio between the two sides should not be parametrically large, even
if one scales all the $\zeta_i$ by a common large factor (\ie no new
large ratios should appear in the problem, because they could lead to
large logarithms). This suggests the following more precise condition,
\begin{equation}
  \label{eq:rIRC1}
  \lim_{\bar v \to 0} \frac{V(\kappa_1(\bar v
    \zeta_1),\ldots,\kappa_n(\bar v \zeta_n))}{\bar v}  
   =
  f_{\{\kappa_i\}}(\zeta_1,\ldots,\zeta_n)\,,
\end{equation}
where $f_{\{\kappa_i\}}$ should be some well-defined non-zero function
and where the subscript indicates that it may depend also on the
additional characteristics of the $\kappa_i$ ($\ell_i$, $\xi_i$,
$\phi_i$).

The $\kappa_i$ are also useful in formulating the precise conditions
needed to go from fig.\ref{fig:clusters}a to fig.\ref{fig:clusters}b.
For example we needed the observable to be such that we could neglect
the grey emissions, leaving only those in the band between the dashed
and dotted lines --- furthermore we needed to be able to require that
the width of the band, $\ln 1/\epsilon$, was not parametrically large,
\ie would not be associated with extra logarithms. This is equivalent
to saying that in eq.~(\ref{eq:rIRC1}), independently of $\bar v$ on
the left-hand-side, there should exist an $\epsilon$ such that any
emissions with $\zeta_i < \epsilon$ can be neglected, or
\begin{equation}
  \label{eq:rIRC2}
  \lim_{\zeta_n \to 0} f_{\{\kappa_i\}}(\zeta_1,\ldots,\zeta_n) = 
  f_{\{\kappa_i\}}(\zeta_1,\ldots,\zeta_{n-1})
  \,.
\end{equation}
This looks very much like infrared safety, but is stronger because
$f_{\{\kappa_i\}}$ already involves an infrared limit of its own.
Accordingly we call it \emph{recursive} infrared safety. Making use
simultaneously of normal infrared safety it can also be rewritten in
terms of a commutator of limits,
\begin{equation}
  \label{eq:rIRC2alt}
  \left[\lim_{\zeta_n \to 0},
    \lim_{\bar v \to 0}\right]
  \frac{V(\kappa_1(\bar v
    \zeta_1),\ldots,\kappa_n(\bar v \zeta_n))}{\bar v}  
   =
   0\,.
\end{equation}
Similarly, to ensure that one can replace the clusters of black
emission in fig.\ref{fig:clusters}a with single emissions in
fig.\ref{fig:clusters}b, one needs a condition called recursive
collinear safety.

Given these recursive IRC (rIRC) conditions, exponentiation is
guaranteed, and to NLL accuracy one has
\begin{multline}
  \label{eq:lnSigma-res}
  \ln \Sigma_V(v) = L {\cal G}_1(\as L, a, \{b_\ell\}) +\\
                    + {\cal G}_2(\as L, a, \{b_\ell\}, \{\bar d_\ell\})
                    + {\cal F}_V(R')\,,
\end{multline}
where the ${\cal G}_i$ are known analytical functions whose observable
dependence enters only through the $a$, $b_\ell$ and $\bar d_\ell$
coefficients; $R' = \partial_{L} (L{\cal G}_1(\as L,a,\{b_\ell\}))$
and the NLL ${\cal F}_V(R')$ accounts for the observable's non-trivial
dependence on multiple emissions. In most cases it is given by
\begin{multline}
  \label{eq:cF}
  {\cal F}_V(R') = \lim_{\epsilon\to0}   \frac{\epsilon^{R'}}{R'}
    \sum_{m=0}^{\infty} \frac{1}{m!}
  \times \\ \times
  \left( \prod_{i=1}^{m+1} \sum_{\ell_i=1}^n  R_{\ell_i}'
    \int_{\epsilon}^{1} \frac{d\zeta_i}{\zeta_i} 
    \int_0^{2\pi} \frac{d\phi_i}{2\pi}
  \right)   \delta\!\left(\ln \zeta_1\right)
  \times \\ \times
  \exp\left(-R'\ln
    f_{\{\kappa_i\}}(\zeta_1,\ldots,\zeta_{m+1})\right)\,.
\end{multline}
where $R_{\ell_i}'$ is the contribution to $R'$ from leg $\ell$.

Resummation is automated through an expert system (CAESAR
\cite{autosum}) which uses high precision arithmetic to (a) probe the
observable in the soft and collinear region to obtain the $a$,
$b_\ell$ and $d_\ell$ coefficients and the $g_\ell(\phi)$, (b) test
the rIRC safety, (c) evaluate $\cF_V(R')$ via Monte Carlo integration.

%----------------------------------------------------------------------
%\section{PP DIJET EVENT SHAPES}

One of the most non-trivial applications of CAESAR is to dijet event
shapes at proton-(anti)proton colliders. Such event shapes are of
interest for various reasons, among them sensitivity to a new class of
soft colour-evolution anomalous dimensions \cite{ColourEv}, and the
prospect of extracting information about non-perturbative correction
from gluon hadronisation and from the underlying event. Unfortunately
the event shapes measured so far \cite{TevEvShp} are not global and
therefore not within the scope of CAESAR. Globalness is a significant
issue at hadron-colliders because detectors are not able to measure
the whole event, but only up to some limited maximum rapidity, $\eta_{\max}
\simeq 3.5$ at the Tevatron ($\sim 5$ for LHC).

There are two main ways of working around this problem. Firstly one
can define an event shape that measures everywhere (directly global)
\cite{Banfi:2001ci}, and then note that if $\eta_{\max}$ is sufficiently
large, then the error introduced by actually measuring only particles
with $|\eta| < \eta_{\max}$ is of order 
%\begin{equation}
%  \label{eq:DeltaV}
$\delta V = {\bar d}_i e^{-(a+b_{i}) \eta_{\max}}, $
%\end{equation}
%
where $b_{i}$, $\bar d_i$ are the $b_\ell$, $\bar d_\ell$ coefficients
for the incoming hadron legs. As long as most of the event shape
distribution is concentrated at values of the observable much larger
than this, then the non-measurement of particles with
$|\eta|>\eta_{\max}$ has a negligible impact on the distribution.

\begin{table*}[t]
  \caption{Properties of the various event shapes proposed in
    \cite{hhresum}. Entries marked with an asterisk are subject to
    uncertainties.}
  \centering
  \newcommand\TT{\rule{0pt}{3.3ex}}
  \newcommand\BB{\rule[-2.2ex]{0pt}{0pt}}
  {
  \begin{tabular}[t]{|c|c|c|c|c|} \hline
    Event-shape & Impact of $\eta_{\max}$ & \TT\BB
    \begin{minipage}[c]{0.18\tw}
      \begin{center}
        Resummation breakdown
      \end{center}
    \end{minipage}
    & 
    \begin{minipage}[c]{0.15\tw}
      \begin{center}
        Underlying Event
      \end{center}
    \end{minipage}
    & 
    \begin{minipage}[c]{0.17\tw}
      \begin{center}
        Jet hadronisation
      \end{center}
    \end{minipage}\\ \hline\hline
    %%%%%%%%%%%
    $\tau_{\perp,g}$ & tolerable$^{*}$ & none & $\sim \eta_{\max}/Q$ & $\sim 1/Q$
    \\ \hline
    $T_{m,g}$ & tolerable & none& $\sim \eta_{\max}/Q$ & $\sim 1/(\sqrt{\as} Q)$
    \\ \hline
    $y_{23}$ & tolerable & none& $\sim \sqrt{y_{23}}/Q^*$ & $\sim \sqrt{y_{23}}/Q\, ^*$
    \\ \hline\hline
    $\tau_{\perp,\cE}$, $\rho_{X,\cE}$ & negligible & none& $\sim 1/Q$ & $\sim 1/Q$
    \\ \hline
    $B_{X,\cE}$ & negligible & none& $\sim 1/Q$ & $\sim
    1/(\sqrt{\as} Q)$
     \\ \hline\hline
    $T_{m,\cE}$ & negligible & serious& $\sim 1/Q$ & $\sim 1/(\sqrt{\as}Q)$
    \\ \hline
    $y_{23,\cE}$ & negligible & none& $\sim 1/Q$ & $\sim \sqrt{y_{23}}/Q \,^{*}$
    \\ \hline\hline
    $\tau_{\perp,\cR}$, $\rho_{X,\cR}$ & none & serious & $\sim 1/Q$ & $\sim 1/Q$
    \\ \hline
    $T_{m,\cR}$, $B_{X,\cR}$ & none & tolerable  & $\sim 1/Q$ & $\sim
    1/(\sqrt{\as} Q)$
    \\ \hline
    $y_{23,\cR}$ & none & intermediate$^{*}$ &  $\sim \sqrt{y_{23}}/Q \,^*$ & $\sim \sqrt{y_{23}}/Q \, ^*$
    \\ \hline
  \end{tabular}
  }
  \label{tab:hhevshp}
\end{table*}

\begin{figure}[htbp]
  \vspace{-1.5em}
  \centering
  \includegraphics[width=0.8\columnwidth]{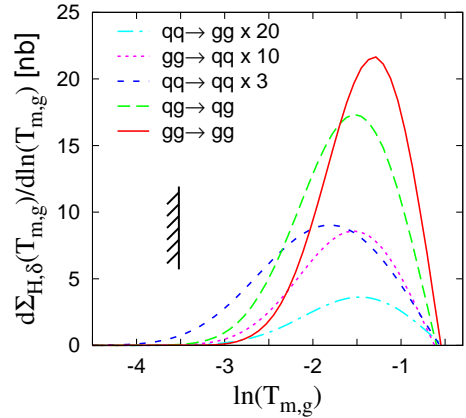}\vspace{-3em}
  \caption{Resummed distribution of $T_{m,g}$ for Tevatron events
    where the hardest jet has $E_\perp > 50\,\mathrm{GeV}$.  The point where
    the $\eta_{\max}=3.5$ cutoff becomes relevant is indicated by a
    hashed line.  \vspace{-2.5em}}
  \label{fig:tmin_HH_dir}
\end{figure}

Among these directly-global observables, there are two subclasses.
Firstly there are those naturally defined in the whole event, a
transverse thrust $\tau_{\perp,g}$, its minor $T_{m,g}$ and the
resolution parameter at which the $k_t$ jet-finder goes from
classifying an event as $2+2$-jet like to $3+2$-jet like, $y_{23}$.
The resummed distribution for $T_{m,g}$ is shown in
fig.~\ref{fig:tmin_HH_dir}.

Alternatively one can define a main event shape in a central region
$\cal C$ (\eg $|\eta| < 1$) and add to it an exponentially-suppressed
contribution for particles in the forward region ${\cal E}_{\cal \bar
  C} = \sum_{i \not\in \cal C} k_{ti}/Q_{\perp,\cal C} e^{-|\eta_i -
  \eta_{\cal C}|},$ where $Q_{\perp,\cal C}$ is the sum of transverse
momenta of particles in $\cal C$ and $\eta_{\cal C}$ is the
(transverse-momentum weighted) mean rapidity of particles in $\cal C$.
Such `combined' observables can be constructed starting from the
`naturally global' event shapes given above, and also from event
shapes that only make sense when defined in the central region such as
jet broadenings and invariant masses, giving \eg $B_{T,{\cal E}}$. The
naturally global observables always have $b_i = 0$ and are more
sensitive to the $\eta_{\max}$ cut (\cnf fig.~\ref{fig:tmin_HH_dir})
than those combined with the forward exponential suppression, which
have $b_i = a$.

One can also define `indirectly-global' observables. Here too there is
a main event shape in the central region and to it one adds a recoil
term ${\cal R}_{\perp \cal C} = \left| \sum_{i \in \cal C} {\vec
    k}_{ti} \right|/Q_{\perp,\cal C}$ measured in $\cal C$, which
through momentum-conservation is sensitive to emissions outside $\cal
C$. This eliminates the problem of sensitivity to $\eta_{\max}$, however
instead one runs into the issue that the CAESAR resummation is
reliable only as long as the main mechanism by which an observable can
take a small value is by suppression of radiation. If instead it can
have a small value through cancellations between different emissions
(as with a vector recoil) then the resummation breaks down via a
divergence of ${\cal F}(R')$ at some small but finite value of $v$.

Table~\ref{tab:hhevshp} summarises the impact of $\eta_{\max}$ and
resummation breakdown (if any) for a range of observables defined in
\cite{hhresum}. It is gives expectations for the degree of sensitivity
to the underlying event and jet hadronisation, illustrating the
complementarity between observables.

%----------------------------------------------------------------------
%\section*{ACKNOWLEDGEMENTS}

The work presented here was carried out in collaboration with A.~Banfi
and G.~Zanderighi. I wish to thank the organisers of the RADCOR 2005
Symposium for the welcoming and stimulating atmosphere of the
conference as well as for financial support.

%======================================================================


\begin{thebibliography}{99}

\bibitem{DSReview}
  M.~Dasgupta and G.~P.~Salam,
  %``Event shapes in e+ e- annihilation and deep inelastic scattering,''
  J.\ Phys.\ G {\bf 30} (2004) R143.
  %[arXiv:hep-ph/0312283].
  %%CITATION = HEP-PH 0312283;%%
  %%Cited 23 times in SPIRES-HEP


\bibitem{NLOJET}
Z.~Nagy,
%``Three-jet cross sections in hadron hadron collisions at next-to-leading  order,''
Phys.\ Rev.\ Lett.\  {\bf 88}, 122003 (2002)
%\prl{88}{2002}{122003}
%[hep-ph/0110315] 
and references therein.
%%CITATION = HEP-PH 0110315;%%

\bibitem{CTTW}
%S.~Catani, L.~Trentadue, G.~Turnock and B.~R.~Webber,
S.~Catani {\it et al.},
%``Resummation of large logarithms in e+ e- event shape distributions,''
Nucl.\ Phys.\ B {\bf 407}, 3 (1993).
%\npb{407}{1993}{3}.
%%CITATION = NUPHA,B407,3;%%

\bibitem{DasSalGlob} 
M.~Dasgupta and G.~P.~Salam,
%``Resummation of non-global QCD observables,''
Phys.\ Lett.\ B {\bf 512}, 323 (2001);
%\plb{512}{2001}{323} 
%[hep-ph/0104277];
%%CITATION = HEP-PH 0104277;%%
%M.~Dasgupta and G.~P.~Salam,
%``Accounting for coherence in interjet E(t) flow: A case study,''
JHEP {\bf 0203} (2002) 017
%\jhep{03}{2002}{017}
%[hep-ph/0203009].
%%CITATION = HEP-PH 0203009;%%


\bibitem{DasSalDIS}
M.~Dasgupta and G.~P.~Salam,
%``Resummed event-shape variables in DIS,''
JHEP {\bf 0208}, 032 (2002).
%\jhep{08}{2002}{032}
%[hep-ph/0208073].
%%CITATION = HEP-PH 0208073;%%

\bibitem{autosum}
  A.~Banfi, G.~P.~Salam and G.~Zanderighi,
  %``Generalized resummation of QCD final-state observables,''
  Phys.\ Lett.\ B {\bf 584} (2004) 298;
  %[arXiv:hep-ph/0304148];
  %%CITATION = HEP-PH 0304148;%%
  %%Cited 24 times in SPIRES-HEP
  %``Principles of general final-state resummation and automated
  %implementation,''
  JHEP {\bf 0503} (2005) 073.
  %[arXiv:hep-ph/0407286]; \url{http://qcd-caesar.org/}
  %%CITATION = HEP-PH 0407286;%%
  %%Cited 19 times in SPIRES-HEP


\bibitem{ColourEv}
J.~Botts and G.~Sterman,
%``Hard Elastic Scattering In QCD: Leading Behavior,''
Nucl.\ Phys.\ B {\bf 325}, 62 (1989);
%\npb{325}{1989}{62}.
%%CITATION = NUPHA,B325,62;%%
N.~Kidonakis and G.~Sterman,
%``Subleading logarithms in QCD hard scattering,''
Phys.\ Lett.\ B {\bf 387} (1996) 867;
%\plb{387}{1996}{867};
N.~Kidonakis, G.~Oderda and G.~Sterman,
%``Evolution of color exchange in {QCD} hard scattering,''
Nucl.\ Phys.\ B {\bf 531}, 365 (1998);
%\npb{531}{1998}{365} 
%[hep-ph/9803241]. 
%
%R.~Bonciani, S.~Catani, M.~L.~Mangano and P.~Nason,
R.~Bonciani {\it et al.},
%``Sudakov resummation of multiparton QCD cross sections,''
Phys.\ Lett.\ B {\bf 575} (2003) 268.
%\plb{575}{2003}{268}
%[hep-ph/0307035].
%%CITATION = HEP-PH 0307035;%%


\bibitem{TevEvShp}
F.~Abe {\it et al.}  [CDF Collaboration],
 %``Measurement of QCD jet broadening in p anti-p collisions at S**(1/2) =
%1.8-TeV,''
Phys.\ Rev.\ D {\bf 44} (1991) 601;
%\prd{44}{1991}{601}.
%%CITATION = PHRVA,D44,601;%%
%
%\bibitem{D0Thrust}
I.~A.~Bertram  [D0 Collaboration],
%``Jet Results At The D0 Experiment,''
Acta Phys.\ Polon.\ B {\bf 33}, 3141 (2002).
%\appol{B 33}{2002}{3141}.
%%CITATION = APPOA,B33,3141;%%

\bibitem{Banfi:2001ci}
  %A.~Banfi, G.~Marchesini, G.~Smye and G.~Zanderighi,
  A.~Banfi {\it et al.},
  %``Out-of-plane QCD radiation in DIS with high p(t) jets,''
  JHEP {\bf 0111}, 066 (2001).
  %[arXiv:hep-ph/0111157].
  %%CITATION = HEP-PH 0111157;%%
  %%Cited 20 times in SPIRES-HEP

\bibitem{hhresum}
  A.~Banfi, G.~P.~Salam and G.~Zanderighi,
  %``Resummed event shapes at hadron hadron colliders,''
  JHEP {\bf 0408} (2004) 062.
  %[arXiv:hep-ph/0407287].
  %%CITATION = HEP-PH 0407287;%%
  %%Cited 13 times in SPIRES-HEP



\end{thebibliography}
\end{document}